\documentclass[conference]{IEEEtran}
\IEEEoverridecommandlockouts
\usepackage{cite}
\usepackage{amsmath,amssymb,amsfonts}
\usepackage{algorithmic}
\usepackage{graphicx}
\usepackage{textcomp}
\usepackage{xcolor}
\usepackage{hyperref}
\usepackage{enumitem}
\usepackage{graphicx} 
\usepackage{array} 

\usepackage{booktabs} 
\def\BibTeX{{\rm B\kern-.05em{\sc i\kern-.025em b}\kern-.08em
    T\kern-.1667em\lower.7ex\hbox{E}\kern-.125emX}}

\makeatletter
 \let\old@ps@headings\ps@headings
 \let\old@ps@IEEEtitlepagestyle\ps@IEEEtitlepagestyle
 
 \def\confheader#1{%
  \def\ps@IEEEtitlepagestyle{%
    \old@ps@IEEEtitlepagestyle%
    \def\@oddhead{\strut\hfill#1\hfill\strut}%
    \def\@evenhead{}%
  }%
  \def\ps@headings{%
    \old@ps@headings%
    \def\@oddhead{}%
    \def\@evenhead{}%
  }%
 \ps@headings%
 }

 \makeatother

\confheader{Accepted in 2025 IEEE International Conference on Service-Oriented System Engineering (SOSE)}

\usepackage[pscoord]{eso-pic}
\newcommand{\placetextbox}[3]{
 \setbox0=\hbox{#3}
 \AddToShipoutPictureFG*{ \put(\LenToUnit{#1\paperwidth},\LenToUnit{#2\paperheight}){\vtop{{\null}\makebox[0pt][c]{#3}}}
 }
 }
 \placetextbox{.21}{0.055}{\small{979-8-3315-0769-5/25/\$31.00~\copyright 2025 IEEE}}

\begin{document}

\title{Bridging Cloud Convenience and Protocol Transparency: A Hybrid Architecture for Ethereum Node Operations on Amazon Managed Blockchain}

\author{
    \IEEEauthorblockN{S M Mostaq Hossain, Amani Altarawneh and Maanak Gupta}
    \IEEEauthorblockA{\textit{Department of Computer Science}, \textit{Tennessee Technological University}\\
    Cookeville, Tennessee, USA \\
    Email: \{shossain42, aaltarawneh, mgupta\}@tntech.edu}
}

\maketitle

\begin{abstract}
As blockchain technologies are increasingly adopted in enterprise and research domains, the need for secure, scalable, and performance-transparent node infrastructure has become critical. While self-hosted Ethereum nodes offer operational control, they often lack elasticity and require complex maintenance. This paper presents a hybrid, service-oriented architecture for deploying and monitoring Ethereum full nodes using Amazon Managed Blockchain (AMB), integrated with EC2-based observability, IAM-enforced security policies, and reproducible automation via the AWS Cloud Development Kit. Our architecture supports end-to-end observability through custom EC2 scripts leveraging Web3.py and JSON-RPC, collecting over 1,000 real-time data points—including gas utilization, transaction inclusion latency, and mempool dynamics. These metrics are visualized and monitored through AWS CloudWatch, enabling service-level performance tracking and anomaly detection. This cloud-native framework restores low-level observability lost in managed environments while maintaining the operational simplicity of managed services. By bridging the simplicity of AMB with the transparency required for protocol research and enterprise monitoring, this work delivers one of the first reproducible, performance-instrumented Ethereum deployments on AMB. The proposed hybrid architecture enables secure, observable, and reproducible Ethereum node operations in cloud environments, suitable for both research and production use.
\end{abstract}



\section{Introduction} \label{sec:introduction}
Blockchain technology has transformed the landscape of secure, decentralized computation and data management across domains such as finance, supply chain, and critical infrastructure systems~\cite{lin2022survey}. Ethereum, as a widely adopted smart contract platform, plays a pivotal role in enabling trustless interactions and decentralized applications (DApps)~\cite{zheng2023blockchain}. The ability to deploy and operate Ethereum nodes~\cite{zhang2022ethereum} efficiently and securely is essential not only for application developers and enterprises, but also for researchers and regulators studying network behavior, performance, and resilience.

Traditionally, running an Ethereum node~\cite{prysmInstall2024} requires setting up a Geth or Besu client on self-hosted infrastructure, involving ongoing maintenance, patching, peer configuration, and resource provisioning. While this provides flexibility, it also introduces significant operational overhead and security risks—especially when misconfigured. As adoption grows, enterprises increasingly turn to managed solutions~\cite{kolb2020core} that simplify operations without compromising security, observability, or compliance. To meet this demand, Amazon Web Services (AWS)~\cite{mathew2014overview} offers Amazon Managed Blockchain~\cite{awsAMB2024}—a fully managed service for deploying and managing blockchain nodes and configurations using frameworks like Hyperledger Fabric and Ethereum~\cite{jayadev2024assessing}. AMB handles infrastructure, networking, and updates, streamlining Ethereum node deployment. However, its trade-offs, performance, and operational implications remain underexplored in academic research.

Despite the convenience of managed services~\cite{khan2022blockchain}, several key aspects remain underexplored. The performance of AMB nodes under varying workloads has yet to be thoroughly benchmarked. Security mechanisms—particularly for protecting against endpoint exposure, peer churn, and denial-of-service attacks~\cite{ibrahim2022ddos}—also require further validation. Moreover, managed deployments lack support for empirical protocol analysis, gas-based prioritization studies, and fine-grained observability, especially on public mainnets. Limited visibility into node behavior and mempool dynamics~\cite{riedelMempool} poses challenges for researchers seeking operational transparency. These limitations highlight the need for hybrid architectures that combine the simplicity of managed services with external observability and security instrumentation. Existing literature has largely focused on self-hosted blockchain deployments or theoretical analyses of Ethereum’s protocol behavior~\cite{loghin2025characterizing}. Limited empirical research exists on cloud-managed Ethereum deployments~\cite{loghin2022blockchain}, especially in enterprise contexts where compliance, uptime, and observability are paramount. This creates a knowledge gap for both practitioners and researchers aiming to assess managed blockchain services in production or academic experiments.

This paper addresses the lack of transparent, reproducible performance evaluations for Ethereum nodes on managed blockchain platforms by introducing a comprehensive framework for deploying, monitoring, and analyzing Amazon Managed Blockchain Ethereum nodes. Data and scripts supporting this work are available at GitHub \footnote{\href{https://github.com/MostaqHossain/aws-amb-eth-ec2}{{https://github.com/MostaqHossain/aws-amb-eth-ec2}}}. The main contributions of this work are:


\begin{enumerate}[label=(\roman*)]
    \item \textbf{Secure and Scalable Node Provisioning:} We implement an AMB-based architecture with AWS Identity and Access Management (IAM)-based access control~\cite{awsIAM2024}, private endpoint isolation, and TLS-encrypted communication to ensure secure infrastructure operation.
    
    \item \textbf{Hybrid Monitoring and Observability:} We develop a custom monitoring layer using EC2 instances~\cite{awsEC22024} and JSON-RPC~\cite{jsonrpc2024} to collect 1,000+ time-series datapoints at 60-second intervals. Observability is enhanced via CloudWatch~\cite{awsCloudWatch2024} dashboards and alerts for latency, throughput, and resource anomalies.
    
    \item \textbf{Automated Infrastructure Deployment:} We use AWS Cloud Development Kit (CDK)~\cite{awsCDK2024} to codify the entire deployment stack, including AMB nodes, EC2 agents, IAM policies, Virtual Private Cloud (VPC)~\cite{awsVPC2024} settings, and CloudWatch integration—ensuring reproducibility across experiments and regions.
    
    \item \textbf{Empirical Blockchain Performance Evaluation:} Using the EC2-based monitoring agents and secure RPC endpoints, we quantify transaction inclusion latency, gas efficiency, mempool clearance behavior, and ETH transfer volume under varying network conditions. These are visualized using high-resolution plots derived from real-time blockchain activity.
    
    \item \textbf{Cost-Aware, Research-Ready Architecture:} By combining managed blockchain services with open monitoring and automation tooling, we demonstrate a secure, cost-effective, and repeatable architecture that can support both enterprise deployments and blockchain performance research.
\end{enumerate}
The rest of this paper is structured as follows:
Section II presents background information on Ethereum and Amazon Managed Blockchain.
Section III reviews related work in Ethereum deployment, performance benchmarking, and blockchain orchestration. 
Section IV describes the system architecture, including design choices and AWS service integration.
Section V outlines our experimental methodology and implementation. Section VI discusses performance results and security analysis.
Section VII identifies limitations and proposes future extensions. Finally, Section VIII concludes the paper.

\section{Background} \label{sec:background}

Ethereum is a decentralized, Turing-complete~\cite{wright2019agent} blockchain platform designed for executing smart contracts. It provides a programmable environment where developers can deploy decentralized applications on a trustless, permissionless~\cite{peng2021privacy} network. Ethereum operates on a proof-of-stake (PoS)~\cite{saleh2021blockchain} consensus mechanism (since the transition in The Merge), replacing the energy-intensive proof-of-work (PoW)~\cite{gervais2016security} model. 
\subsection{Ethereum Blockchain Overview}
Ethereum has several key architectural components, which include:
\begin{itemize}
    \item Accounts: Two types exist—Externally Owned Accounts (EOAs), controlled by private keys, and Contract Accounts, controlled by smart contract code~\cite{lin2024measurement}.
    \item Smart Contracts: Immutable code deployed on-chain, capable of managing assets, verifying logic, and interacting with other contracts~\cite{tikhomirov2017ethereum}.
    \item Gas: A transaction fee mechanism used to meter and limit computation. Gas prices fluctuate based on network congestion, impacting transaction inclusion~\cite{li2021gas}.
    \item Nodes: Nodes are critical to maintaining Ethereum’s state, validating transactions, and propagating new blocks across the peer-to-peer network~\cite{ethereumNodesClients2024}.
\end{itemize}
There are several types of Ethereum nodes~\cite{alchemyNodeTypes2024}:
\begin{itemize}
    \item Full Nodes: Store the entire blockchain history and validate blocks and transactions independently.
    \item Archive Nodes: Extend full node functionality by retaining all historical states for querying and analysis.
    \item Light Clients: Rely on full nodes for data and do not store full blockchain history.
\end{itemize}
The performance, reliability, and visibility of an Ethereum node are heavily influenced by its deployment environment, networking setup, and monitoring capabilities.

\subsection{Challenges of Self-Hosting Ethereum Nodes}
Running a self-hosted Ethereum node—especially in production—presents several operational challenges. First of all, high storage and compute requirements (particularly for full and archive nodes). Then ongoing maintenance, including version upgrades, dependency management, and system hardening. The exposure to security threats such as DDoS attacks, RPC endpoint abuse, and peer churn are also there~\cite{getblockNodes2024}. Additionally, complex network configuration to ensure optimal peer connectivity and timely block propagation. These challenges often act as barriers for researchers, developers, and enterprises seeking to maintain reliable Ethereum infrastructure.
\begin{table*}[ht]
\centering
\caption{Key Features of AMB Ethereum Nodes}
\begin{tabular}{@{}p{3.5cm}p{6.8cm}p{6.8cm}@{}}
\toprule
\textbf{Feature} & \textbf{Description} \\
\midrule
Managed Infrastructure & AWS handles node setup, peer discovery, updates, and fault recovery. \\
High Availability & Nodes are deployed across multiple Availability Zones to ensure fault tolerance. \\
VPC Integration & Nodes can reside within a Virtual Private Cloud for restricted network access. \\
Security & TLS encryption is enforced and access is managed via AWS IAM. \\
Scalability & Supports multiple node types and auto-scaling of concurrent RPC requests. \\
\bottomrule
\end{tabular}
\label{tab:amb_features}
\end{table*}

\subsection{Amazon Managed Blockchain (AMB)}
Amazon Managed Blockchain is a fully managed AWS service that allows users to deploy and manage blockchain nodes without handling the underlying infrastructure~\cite{awsAMB2024}. It supports two blockchain frameworks: Hyperledger Fabric (for private consortium blockchains)~\cite{androulaki2018hyperledger} and Ethereum (for public blockchain access). AMB for Ethereum enables developers to provision Ethereum full or archive nodes via a simple API or console interface. It also connects to the Ethereum mainnet or testnets (e.g., Sepolia~\cite{sepoliaTestnet2024}).
They interact with the blockchain through a secure HTTPS RPC endpoint. By leveraging AWS-native services like IAM, CloudWatch, and VPC for access control and monitoring. The key features of AMB Ethereum nodes are described in Table \ref{tab:amb_features}. Despite these advantages, AMB also introduces abstraction trade-offs such as limited visibility into low-level logs or peer configuration. Also, no access to the file system or Geth client parameters. And inflexibility for users needing fine-grained control over node behavior. These limitations motivate the need for hybrid architectures that combine AMB’s managed services with external observability tools.


\subsection{Comparative Context}
Table \ref{tab:node_comparison} presents a concise comparison between self-hosted Ethereum nodes and those deployed using Amazon Managed Blockchain. The comparison highlights differences in operational control, deployment complexity, monitoring capabilities, and suitability for various use cases. This contextual analysis supports the motivation for a hybrid architecture that balances manageability with observability.

\begin{table}[ht]
\centering
\caption{Comparison of Managed vs. Self-Hosted Ethereum Nodes}
\begin{tabular}{p{1.3cm} p{3cm} p{3.2cm}}
\toprule
\textbf{Aspect} & \textbf{Self-Hosted Nodes} & \textbf{AMB-Hosted Nodes} \\
\midrule
Maintenance & Manual setup and updates & Handled by AWS \\
Peers & Fully customizable & Abstracted and managed \\
Logging & Full access to logs & Limited to external metrics \\
Security & User-defined (firewalls, Access Control Lists) & IAM, VPC, and security groups \\
Monitoring & Custom integration required & Integrates with CloudWatch \\
Deployment & Time-consuming and error-prone & Rapid and scriptable via CDK \\
Use-case & Experimentation, custom tuning & Production-grade, secure access \\
\bottomrule
\end{tabular}
\label{tab:node_comparison}
\end{table}

\subsection{Rationale for Hybrid Monitoring}
To compensate for the reduced visibility in AMB deployments, this paper proposes a hybrid monitoring architecture that uses external EC2 instances to collect blockchain metrics from AMB via RPC. This architecture bridges the gap between convenience and control, enabling deeper insights while maintaining the operational benefits of managed infrastructure.


\section{Related Works} \label{sec:related_works}
The adoption and deployment of Ethereum blockchain nodes in cloud environments have been explored from multiple dimensions, including scalability, performance optimization, security, and operational efficiency. However, most existing research focuses on self-hosted or manually optimized deployments, with limited exploration of managed services such as Amazon Managed Blockchain. This section summarizes key contributions in three thematic areas: Ethereum node performance and cost optimization, security and monitoring in cloud-based blockchain networks, and automation and orchestration frameworks for blockchain deployment.

\subsection{Performance and Cost Optimization of Ethereum Nodes}
Previous work has primarily emphasized resource provisioning and throughput optimization in Ethereum networks. Zhang et al.~\cite{zhang2021evaluation} proposed performance models for Ethereum nodes based on transaction latency and CPU utilization under variable workloads. Their findings underscore the need for elastic resource scaling to maintain acceptable performance under high mempool pressure. Complementing this, Li et al.~\cite{li2021gas} analyzed gas usage inefficiencies in Ethereum smart contracts and proposed compiler-level gas optimization techniques to reduce execution overhead. While these works address performance bottlenecks and computation costs, they assume a self-managed deployment model and lack practical evaluations on cloud-native managed services such as AMB. In contrast, AWS documentation~\cite{awsAMB2024} provides benchmark guidance for tuning node types and storage backends on AMB. However, it is vendor-centric and lacks reproducible experimental insights or cross-comparison with non-managed deployments. Our study addresses this gap by empirically evaluating AMB node performance using hybrid monitoring via EC2 instances and AWS CloudWatch.

\begin{figure*}[ht]
    \centering
    \includegraphics[width=0.9\textwidth]{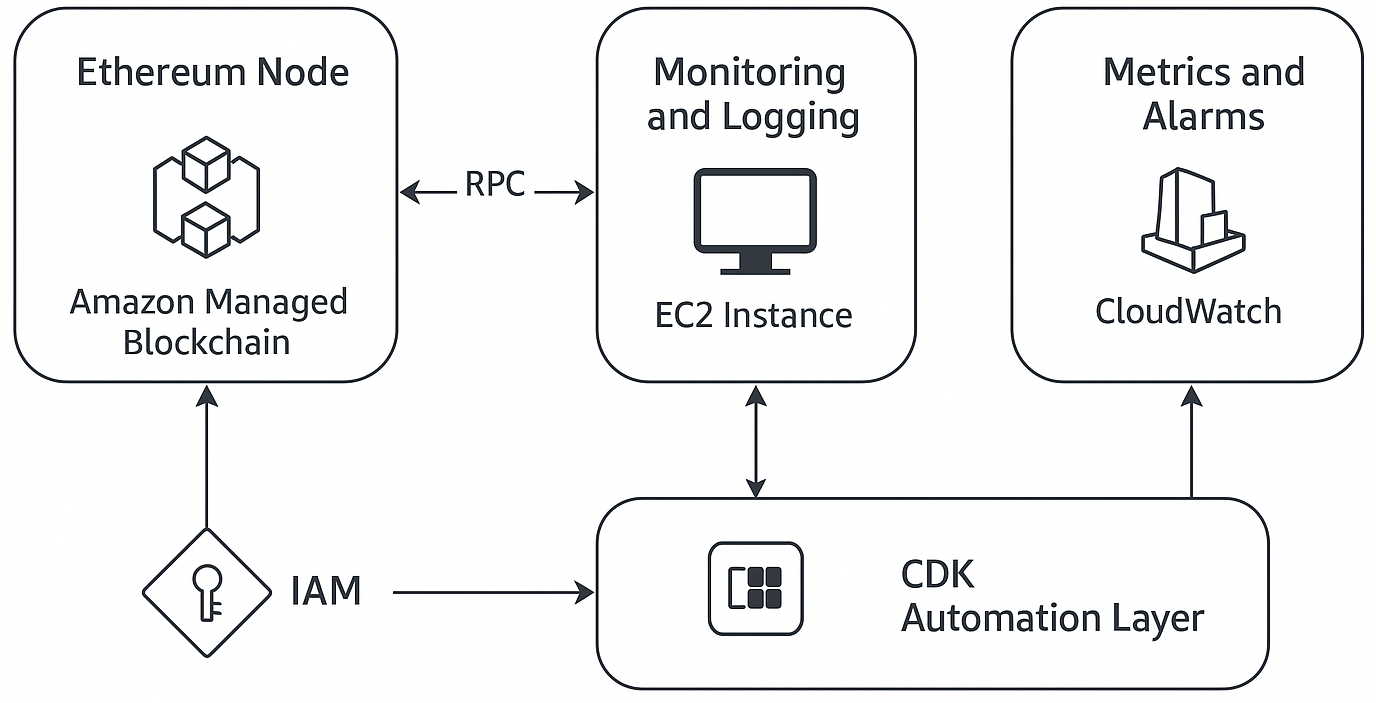}
    \caption{System architecture showing the integration of Amazon Managed Blockchain with EC2-based monitoring, CloudWatch metrics, and AWS CDK automation.}
    \label{fig:system_architecture}
\end{figure*}

\subsection{Blockchain Security and Monitoring in Cloud Environments}
Several researchers have studied the security posture of blockchain networks hosted in public cloud environments. For instance, Li et al.~\cite{li2021strong} introduced threat models for DDoS attacks on Ethereum nodes, highlighting vulnerabilities exposed by open RPC ports and weak authentication policies. Amazon's Well-Architected Blockchain~\cite{leocadio2025aws} offers a security-first approach for deploying blockchain workloads on AWS. It recommends using secure endpoints, VPC isolation, IAM-based access control, and service-level encryption. However, academic assessments of these configurations' actual effectiveness in production-grade Ethereum deployments remain scarce. Moreover, work by Gervais et al.~\cite{gervais2016security} on Ethereum network propagation and consensus delay reveals how suboptimal peer connectivity can degrade security guarantees. Their insights are critical for understanding how cloud-managed networking, such as that in AMB, affects peer discovery and block propagation. Our architecture builds on these principles by implementing secure node endpoints and IAM-based access policies in AMB, while supplementing it with EC2-based observability to detect peer churn, endpoint health, and consensus lag.

\begin{table*}[ht]
\centering
\caption{Comparison of Related Work and This Work}
\label{tab:comparison}
\begin{tabular}{@{}p{3.5cm}p{6.8cm}p{6.8cm}@{}}
\toprule
\textbf{Theme} & \textbf{Related Work} & \textbf{This Work} \\
\midrule
\textbf{Performance \& Cost} & 
Self-managed deployments; theoretical models~\cite{zhang2021evaluation, li2021gas}; AWS AMB docs lack reproducibility. &
Empirical benchmarking of AMB nodes with hybrid monitoring using EC2 and CloudWatch. \\

\textbf{Security \& Monitoring} & 
DDoS threats, RPC exposure~\cite{li2021strong}; peer latency impact~\cite{gervais2016security}; AWS security guidelines. & 
Hardened AMB nodes using IAM, VPC; EC2-based observability for peer churn and consensus delay. \\

\textbf{Automation \& Orchestration} & 
Bevel/Terraform~\cite{hyperledgerBevel} for Fabric and Geth; AWS CDK lacks integrated performance/security support~\cite{leocadio2025aws}. & 
Extended CDK automation with EC2 logging, custom setup scripts, and hybrid observability. \\

\textbf{Research Gap} & 
Academic focus on self-hosted/testnet setups~\cite{lal2021blockchain}; managed services underexplored~\cite{maksymyuk2022blockchain}. & 
First reproducible, security-integrated evaluation of Ethereum on AMB for research and enterprise. \\
\bottomrule
\end{tabular}
\end{table*}

\subsection{Automation and Orchestration of Blockchain Deployments}
Infrastructure-as-Code (IaC) and automation frameworks have been increasingly adopted for blockchain deployment. Projects like Hyperledger Bevel~\cite{hyperledgerBevel} and Terraform modules for Ethereum setup~\cite{scbTerraform2021} demonstrate modular deployment strategies but focus on Fabric and self-managed Geth instances, respectively. AWS CDK has emerged as a versatile tool for deploying cloud-native blockchain stacks. While AWS provides official CDK patterns for AMB, these are limited in scope and do not include integrated security validation, performance benchmarking, or hybrid design strategies. Our implementation extends CDK patterns by introducing automated setup scripts for EC2 integration, custom log ingestion, and continuous performance tracking with CloudWatch. We also propose a hybrid model combining AMB’s fault-tolerant management with EC2’s transparency for research-grade observability.

\subsection{Research Gap and Contribution}
To the best of our knowledge, no academic work has systematically analyzed the deployment, security, and performance characteristics of Ethereum nodes on Amazon Managed Blockchain. Existing studies either focus on theoretical performance metrics, private testnets, or self-hosted environments, without accounting for the design trade-offs offered by managed services. Our work uniquely contributes a reproducible framework for benchmarking Ethereum nodes on AMB. It integrates security best practices, automated deployment via AWS CDK, and monitoring through CloudWatch and EC2. This makes our architecture suitable not only for secure enterprise adoption but also for academic experimentation. Table~\ref{tab:comparison} summarizes the key differences between existing approaches and our work, highlighting how our implementation fills the gap in performance benchmarking, security integration, and automation for Ethereum nodes on Amazon Managed Blockchain.

\section{System Architecture} \label{sec:system_architecture}
The proposed architecture provides a secure, observable, and scalable framework for deploying Ethereum nodes using Amazon Managed Blockchain. It integrates multiple AWS services to enhance visibility, automate provisioning, and enforce security controls. The design aims to strike a balance between the convenience of managed infrastructure and the flexibility required for performance benchmarking and research experimentation.

\subsection{Overview of Architectural Design}
Figure \ref{fig:system_architecture} illustrates the high-level architecture of our system. At its core, the design relies on a managed Ethereum node provisioned via AMB, which serves as the blockchain access point. To overcome the abstraction limitations of AMB, we augment the architecture with external monitoring and automation layers built on EC2, CloudWatch, and AWS CDK. The system comprises five core components:
\begin{itemize}
    \item Ethereum Node via Amazon Managed Blockchain
    \item Monitoring and Logging Layer on EC2
    \item CloudWatch for Metrics Collection and Alarms
    \item AWS CDK for Infrastructure Automation
    \item Security and Identity Management via IAM and VPC
\end{itemize}

\subsection{Ethereum Node Provisioning with AMB}
AMB provides Ethereum nodes as a fully managed service. In the proposed architecture, a full archival node is provisioned using AMB, allowing us to access historical and real-time blockchain data through a secure RPC endpoint. The node is deployed within a dedicated VPC, with endpoint access restricted to a defined set of IP addresses via security groups. The key \textit{advantages} of using AMB include: automatic software updates and patching, resilience and high availability via AWS-managed clustering, simplified maintenance and scaling and TLS-secured RPC interface. However, AMB abstracts low-level configurations such as peer selection, storage backend, and direct access to logs, which necessitates an auxiliary monitoring mechanism.

\subsection{EC2-Based Monitoring Layer}
To enable transparent monitoring, an EC2 instance connects to the AMB-hosted Ethereum node via RPC and runs custom Web3.py scripts to collect blockchain data, including blocks, transactions, gas usage, and mempool activity. Metrics such as RPC latency and block finalization time are logged to CloudWatch for centralized analysis. This hybrid setup restores observability, supporting use cases like throughput analysis, anomaly detection, and performance benchmarking under simulated workloads.

\subsection{Metrics and Alerts with CloudWatch}
Amazon CloudWatch is tightly integrated into the architecture to provide comprehensive observability across the blockchain system. Metrics are collected from two primary sources: the EC2 monitoring scripts and native AMB service outputs (where accessible). Key indicators such as block latency, finalization delay, RPC response success or failure rates, and EC2 instance resource utilization (including CPU and memory) are continuously monitored. Transaction inclusion statistics are also captured to measure end-to-end performance. These metrics are visualized through custom dashboards within CloudWatch, providing a real-time view of system health. Alarms are configured to notify administrators in the event of anomalies such as RPC endpoint unavailability or unexpectedly prolonged block intervals. This monitoring infrastructure establishes a critical link between the managed blockchain environment and customizable analytics, supporting both operational oversight and deeper academic analysis.

\subsection{Infrastructure Automation via AWS CDK}
To ensure repeatability, modularity, and ease of deployment, the entire system is codified using the AWS CDK. CDK templates are developed to define all infrastructure components in code, enabling consistent setup across environments and regions. The templates automate the creation and configuration of AMB Ethereum nodes, the deployment of EC2 instances preloaded with monitoring scripts and IAM roles, and the provisioning of secure VPC networks with subnets and security groups. In addition, the CDK defines CloudWatch log groups, metric filters, and IAM policies enforcing least-privilege access. This infrastructure-as-code approach enables rapid redeployment, consistent security enforcement, and streamlined teardown, significantly reducing manual error and ensuring version-controlled experimental infrastructure suitable for production and research applications.

\subsection{Security Design}
The system architecture adopts a multi-layered security model that aligns with the AWS shared responsibility framework and the best practices recommended in the AWS Well-Architected Framework for Blockchain. Access control is enforced through IAM policies that restrict interaction with AMB and EC2 resources to designated roles and services. Network-level isolation is achieved by deploying the blockchain node within a private VPC, ensuring that it can only be accessed from within secure subnets. Further, security groups and network access control lists (NACL)~\cite{awsNACL} are configured to filter all inbound and outbound traffic based on strict rules. All RPC communication is secured via TLS encryption to prevent eavesdropping and tampering. Optional auditing and threat detection services, such as AWS CloudTrail~\cite{awsCloudTrail} and GuardDuty~\cite{awsGuardDuty}, can be enabled to monitor API usage patterns and detect potential intrusions. These combined measures ensure that both the managed and custom-deployed components of the system are resilient against unauthorized access and network-based attacks.

\section{Experimental Methodology and Implementation} \label{sec:exp_setup}
To evaluate the performance, observability, and deployment efficiency of Ethereum nodes provisioned via AMB, we developed a reproducible experimental framework. This section outlines the methodology for testbed configuration, data collection, transaction injection, monitoring instrumentation, and infrastructure automation.

\subsection{Experimental Objectives}
The primary goals of this experiment are to assess the performance, reliability, and observability of Ethereum nodes deployed on AMB. Specific objectives include measuring transaction throughput and latency under varying load, analyzing gas-price-based prioritization, and monitoring RPC behavior through real-time metrics. The experiments also validate the scalability of EC2-based monitoring and the reproducibility of infrastructure provisioning using AWS CDK.

\subsection{Testbed Configuration}
The experimental testbed was built entirely within the AWS ecosystem, using the following configuration:
\begin{enumerate}[label=(\alph*)]
    \item Ethereum Node: Provisioned via Amazon Managed Blockchain, configured as a full archival node.
    \item Monitoring Instance: A t3.medium Amazon EC2 instance (2 vCPUs, 4 GB RAM) running Ubuntu 22.04 LTS.
    \item Networking: All resources deployed in a private VPC, with public access restricted via security groups.
    \item Automation Tooling: AWS CDK (v2.129.0) used to provision the AMB node, EC2 instance, IAM roles, CloudWatch log groups, and alarms.
\end{enumerate}

\subsection{Data Collection Pipeline}
To analyze node behavior, a set of custom Python scripts was deployed on the EC2 monitoring instance. These scripts used the Web3.py interface to query the Ethereum node via its secure HTTPS RPC endpoint. Key data collected included:
\begin{enumerate}[label=(\alph*)]
    \item Block metadata (block number, size, gas used, timestamp)
    \item Transaction details (gas price, gas limit, inclusion delay)
    \item Mempool tracking (pending transaction count, age)
    \item RPC response times and availability status
\end{enumerate}
Data was polled at 10-second intervals and forwarded to Amazon CloudWatch Logs. Table~\ref{tab:supported_methods} lists the supported JSON-RPC methods successfully used during this process, while Table~\ref{tab:unsupported_methods} highlights unsupported or restricted endpoints—pointing to operational limitations in the managed AMB environment.

\begin{table}[ht]
\centering
\caption{Supported JSON-RPC Methods on AMB Ethereum Node}
\begin{tabular}{ll}
\toprule
\textbf{Method} & \textbf{Message} \\
\midrule
\texttt{web3\_clientVersion} & Success \\
\texttt{eth\_blockNumber} & Success \\
\texttt{eth\_getBlockByNumber} & Success \\
\texttt{eth\_getTransactionByHash} & Success \\
\texttt{eth\_getTransactionReceipt} & Success \\
\texttt{eth\_call} & Success \\
\texttt{eth\_getLogs} & Success \\
\texttt{eth\_gasPrice} & Success \\
\texttt{eth\_estimateGas} & Success \\
\texttt{eth\_getBalance} & Success \\
\texttt{eth\_getCode} & Success \\
\texttt{net\_version} & Success \\
\texttt{net\_listening} & Success \\
\texttt{eth\_syncing} & Success \\
\texttt{eth\_getTransactionCount} & Success \\
\texttt{txpool\_status} & Success \\
\bottomrule
\end{tabular}
\label{tab:supported_methods}
\end{table}

\begin{table}[ht]
\centering
\caption{Unsupported or Restricted JSON-RPC Methods on AMB}
\begin{tabular}{ll}
\toprule
\textbf{Method} & \textbf{Reason} \\
\midrule
\texttt{eth\_sendRawTransaction} & Typed transaction too short \\
\texttt{txpool\_content} & Method not available on AMB \\
\texttt{debug\_traceTransaction} & Restricted method \\
\texttt{eth\_mining} & Not supported by AMB \\
\bottomrule
\end{tabular}
\label{tab:unsupported_methods}
\end{table}

\subsection{Transaction Submission and Data Collection Framework}
To retrieve real-time blockchain data, we developed a custom Python-based data collection pipeline using \texttt{web3.py} and low-level JSON-RPC API calls. The scripts were deployed on a dedicated EC2 instance and configured to interface with the AMB Ethereum node over HTTPS. In addition to EC2-based execution, the same scripts were occasionally run from a local machine for comparative validation of network latency and availability. The collection process included querying block metadata, transaction pool status, gas usage, and transaction confirmation delays. Each request was timestamped and stored in structured log files for subsequent analysis. This approach enabled granular, time-aligned observations of node behavior, RPC responsiveness, and transaction dynamics directly from the blockchain, without requiring internal access to the AMB node infrastructure.

\subsection{CloudWatch-Based Monitoring and Alerts}
Metrics from the EC2 instance were streamed to Amazon CloudWatch, providing dashboards for:
\begin{itemize}
    \item RPC latency trends
    \item Gas price distributions
    \item Block-level throughput and delay
    \item System-level resource usage (CPU, memory, disk I/O)
\end{itemize}
Alarms were configured for anomalous conditions such as high RPC latency or low transaction throughput. These alerts, integrated with Amazon SNS, enabled timely administrative response to potential faults or degradations.

\begin{figure*}[ht]
\centering
\includegraphics[width=\linewidth]{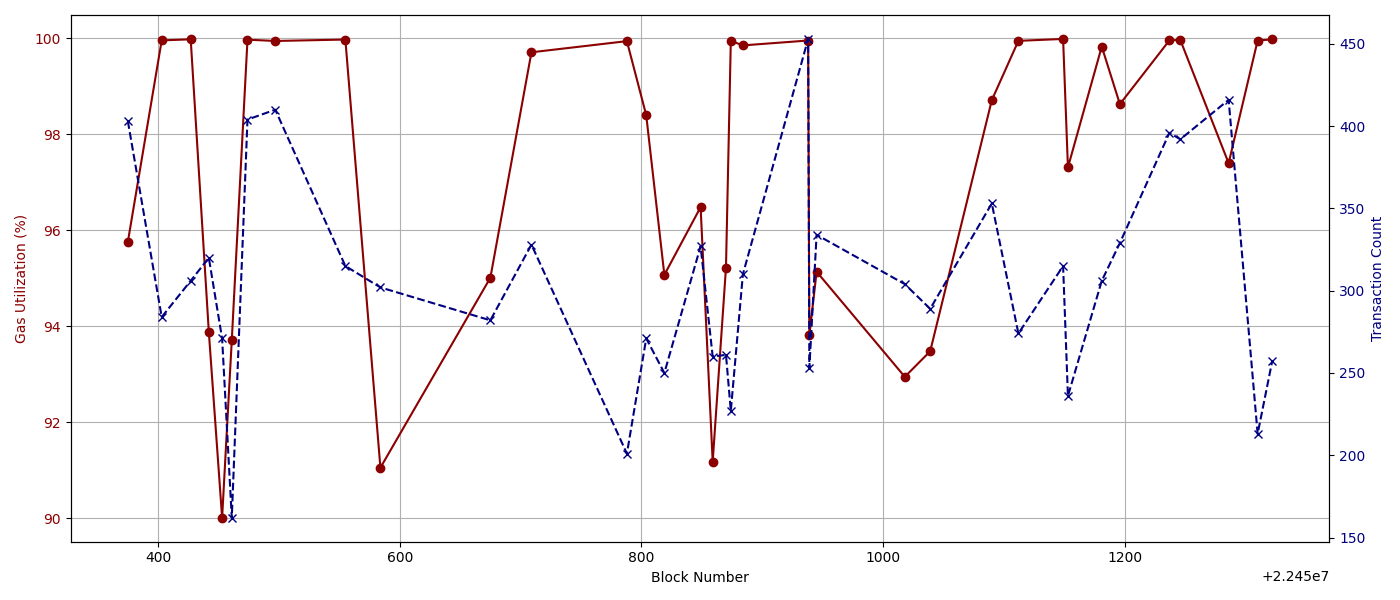} 
\caption{Gas Utilization and Transaction Count for High-Efficiency Blocks. 
This plot focuses exclusively on blocks where gas utilization exceeded 90\%. 
Despite variations in transaction count, these blocks consistently demonstrate optimal use of available gas, 
indicating periods of high on-chain demand or contract-intensive activity.}
\label{fig:high_efficiency_blocks}
\end{figure*}

\subsection{Automation with AWS CDK}
All infrastructure was defined using AWS CDK in TypeScript~\cite{awsCDKTypescript2024}. The CDK project modularized the deployment into reusable stacks, including the AMB node, EC2 instance, IAM roles, VPC settings, and CloudWatch log groups. Post-deployment automation scripts installed dependencies and initialized monitoring tasks on the EC2 instance. This approach ensures version-controlled, reproducible experiments that can be redeployed across regions or adapted for new test cases with minimal overhead.

\subsection{Experiment Duration and Replication}
Each experimental session consisted of collecting 1,000 data points at a fixed 60-second interval per run. This duration was chosen to ensure visibility across multiple block finalization cycles and to capture variations due to Ethereum's dynamic usage patterns, including periods of both low and high transaction volume. To improve the robustness and generalizability of the results, the experiment was replicated across three separate days. All test runs were conducted in the \texttt{us-east-1} (N. Virginia) AWS region to ensure consistent latency profiles and reduce variability caused by geographic differences in blockchain node behavior or AWS infrastructure performance. Aggregated data from each run was statistically analyzed to compute average metrics and identify recurring performance trends, ensuring validity and reproducibility of observations.

\subsection{Deployment Cost Analysis}
To assess operational feasibility, AWS billing was tracked for both blockchain and EC2 services. As summarized in Table~\ref{tab:aws_billing}, the monthly cost breakdown revealed higher expenses for AMB—attributed to continuous node operation and archival data retention—while EC2 costs remained modest, covering monitoring tasks. The three-month total of \$750.29 demonstrates the economic viability of the hybrid setup, balancing observability, performance, and cost-effectiveness for sustained research or enterprise use.

\begin{table}[t]
\scriptsize
\centering
\caption{AWS Billing Summary for Blockchain and EC2 Services (March–May)}
\begin{tabular}{lccc}
\toprule
\textbf{Month} & \textbf{Managed Blockchain (\$)} & \textbf{EC2-Linux (\$)} & \textbf{Total (\$)} \\
\midrule
March & 224.32 & 42.88  & 267.20 \\
April & 336.14 & 72.90 & 409.04 \\
May   & 58.49 & 15.56 & 74.05 \\
\midrule
3 months total    & 618.95 & 131.34 & 750.29 \\
\bottomrule
\end{tabular}
\label{tab:aws_billing}
\end{table}

\section{Performance Results and Security Analysis} \label{sec:discussion}
This section presents empirical results from the deployment and monitoring of an Ethereum node on Amazon Managed Blockchain. Through a series of targeted experiments and data visualizations, we analyze key performance metrics—including gas utilization efficiency, transaction inclusion latency, on-chain economic activity, and mempool behavior—captured via public RPC endpoints and a custom EC2-based monitoring framework. The plots provide quantitative insight into how AMB nodes respond under varying network conditions, illustrating the operational feasibility and performance characteristics of enterprise blockchain deployments in a managed cloud environment. All figures in this section are based on real-time data captured using our EC2-hosted monitoring scripts that queried the AMB Ethereum node's secure RPC endpoint. This approach ensured that the measurements reflect live mainnet activity observed directly through our hybrid telemetry framework, rather than relying on third-party datasets or public blockchain explorers.

\begin{figure*}[!t]
  \centering
  \includegraphics[width=\linewidth, keepaspectratio]{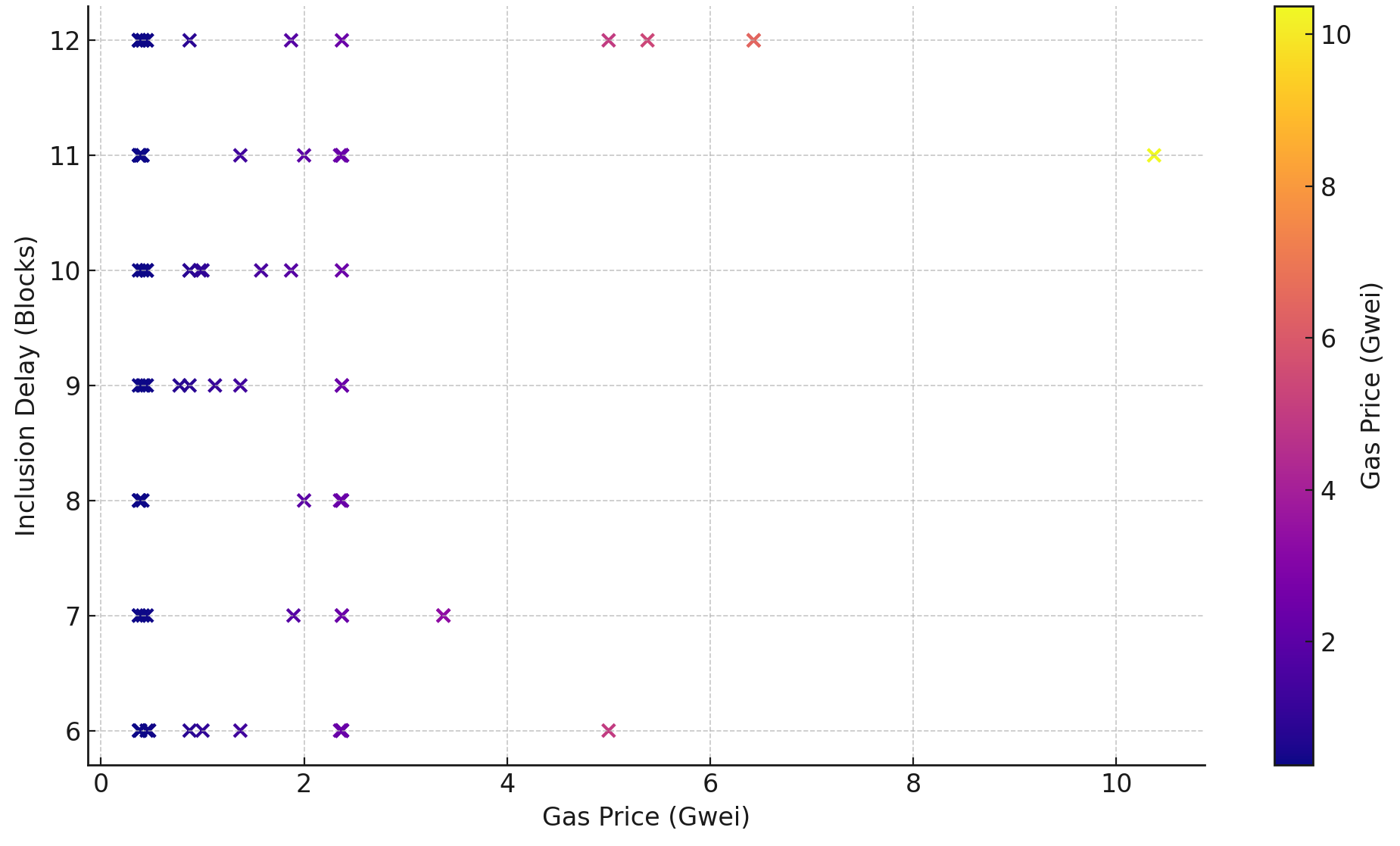}
  \caption{This scatter plot illustrates the relationship between gas price (in Gwei) and transaction inclusion delay (in blocks) using real data from an Ethereum node on Amazon Managed Blockchain. As expected, transactions with higher gas prices are prioritized by validators, resulting in significantly lower confirmation delays—often being included in the very next block—while low-fee transactions experience extended delays due to network congestion and fee market dynamics.}
  \label{fig:tid-gas}
\end{figure*}

\subsection{Finding High Efficiency Blocks}
Figure~\ref{fig:high_efficiency_blocks} presents gas utilization and transaction count for blocks with gas usage above 90\%. The plot reveals that these high-efficiency blocks consistently maximize block capacity, despite variability in transaction count. For example, Block 22450938 processed 453 transactions—the highest recorded—while Block 22450461 handled only 162 transactions yet still exhibited high utilization. This indicates that transaction complexity, rather than count alone, influences gas usage. The figure effectively highlights contract-heavy activity or periods of network congestion where transaction prioritization aligns with gas cost efficiency.

\begin{figure*}[ht]
\centering
\includegraphics[width=\linewidth]{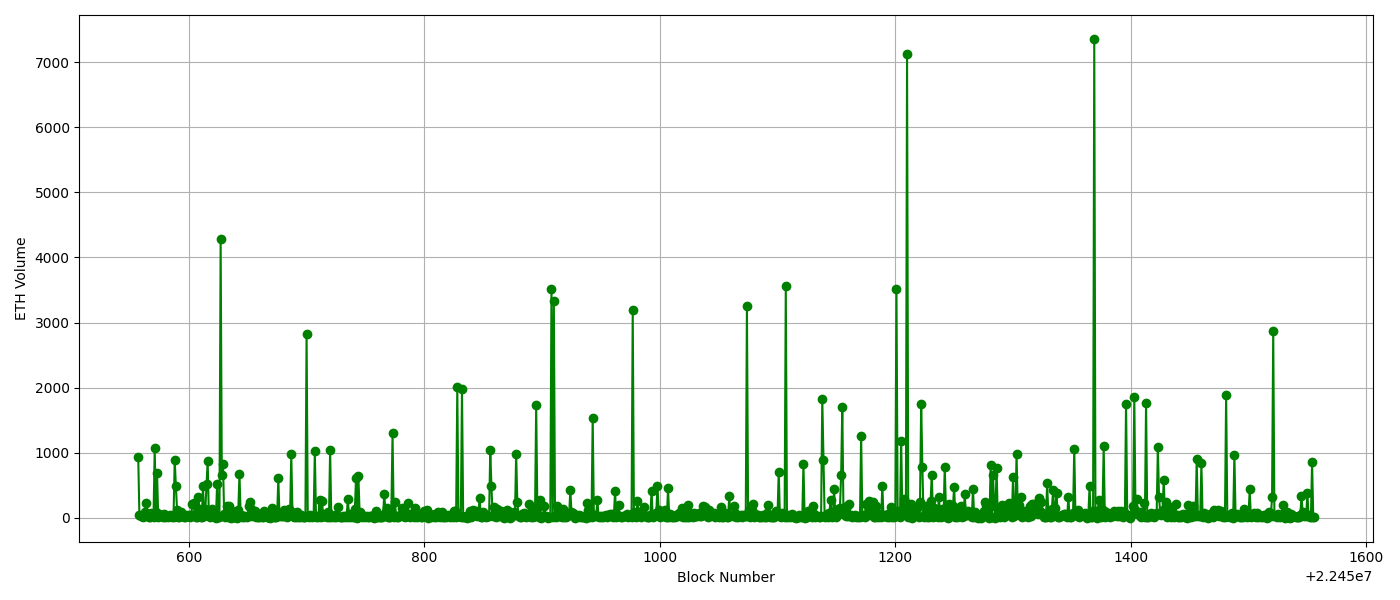}
\caption{Total ETH Transferred Per Block. This plot illustrates the total amount of Ether moved in each block, capturing on-chain economic activity on the AMB Ethereum node. Spikes in volume suggest periods of high-value transfers, possibly due to token swaps, bridge transactions, or large contract settlements, while flatter regions indicate routine or lower-value transactions.}
\label{fig:eth_volume_per_block}
\end{figure*}

\subsection{Transaction Throughput and Latency}
Figure~\ref{fig:tid-gas} illustrates the correlation between gas price and transaction inclusion delay. The figure confirms that higher gas price transactions are prioritized, with many included in the immediate next block, while low-fee transactions endure delays due to network congestion. This empirical relationship reinforces the role of dynamic fee markets in Ethereum and the practical importance of strategic gas bidding for timely execution. The supporting data, derived from randomized gas-priced submissions, provides compelling evidence of AMB node responsiveness and network behavior under varying fee conditions.

\subsection{Total ETH Transferred Per Block}
Figure~\ref{fig:eth_volume_per_block} tracks the volume of Ether transferred per block, showcasing variability in on-chain economic activity. Sharp spikes suggest episodic high-value transfers—possibly driven by large contract interactions, token movements, or batched DeFi transactions—whereas flatter regions indicate routine activity. Notably, this figure emphasizes that high transaction counts do not equate to high ETH volume, underlining the necessity of distinguishing between transaction load and financial throughput in blockchain analytics.


\subsection{Mempool Behavior and Monitoring}
\begin{figure*}[!t]
  \centering
  \includegraphics[width=\linewidth, keepaspectratio]{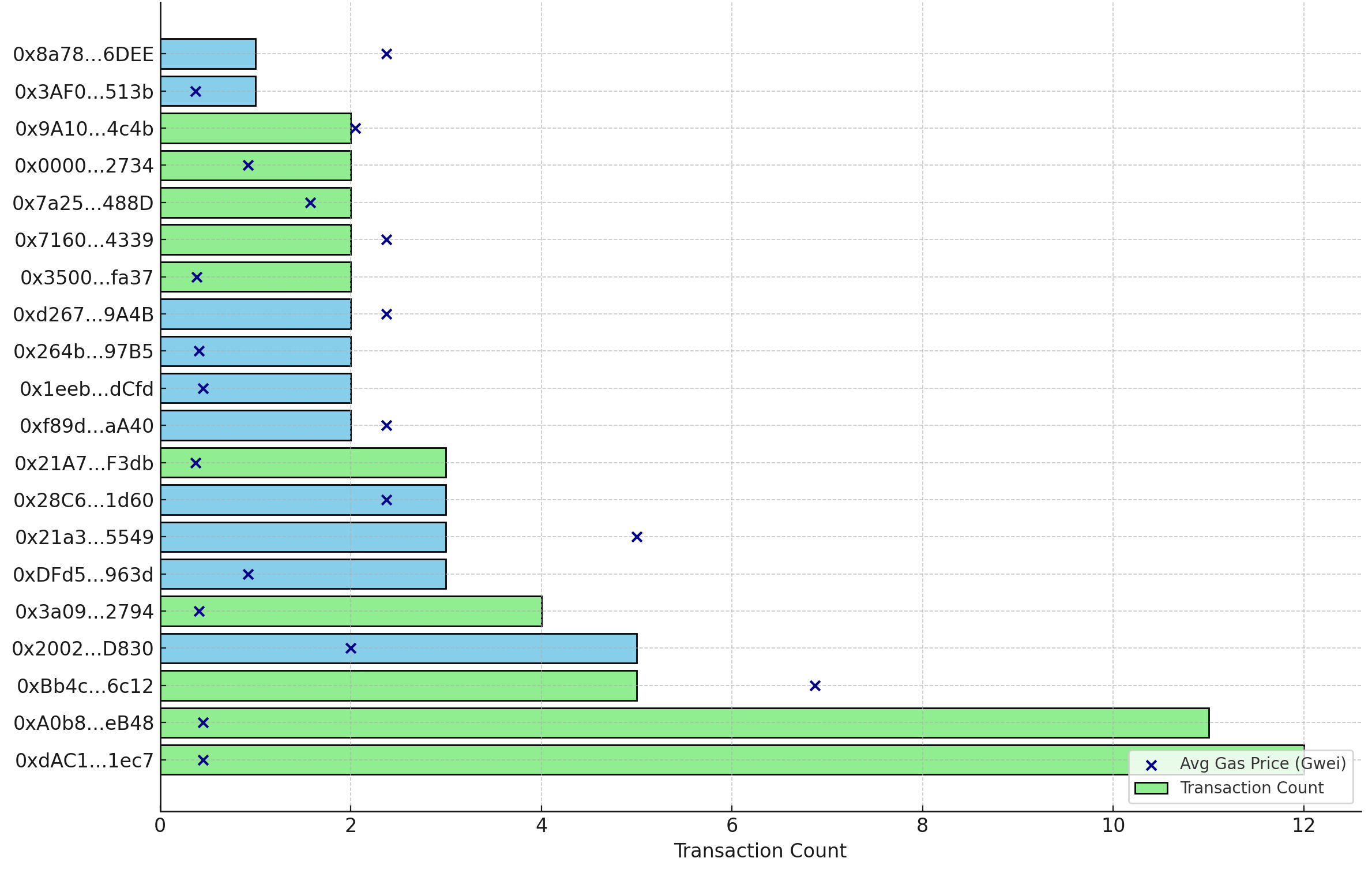}
  \caption{Top Ethereum Addresses by Transaction Count and Average Gas Price. This chart highlights the top 20 sender and receiver addresses based on transaction volume, with overlaid markers showing their average gas price usage, revealing fee strategies among high-activity participants.}
  \label{fig:tc-gas}
\end{figure*}

Figure~\ref{fig:tc-gas} displays the top 20 Ethereum addresses by transaction count, annotated with their average gas price. This visualization highlights the behavior of high-frequency participants and their fee strategies, revealing insights into mempool prioritization under observed network conditions. The combined view of transaction volume and gas expenditure provides a nuanced understanding of how frequent actors navigate the fee market to optimize confirmation speed.

\subsection{CloudWatch Metrics and Alerts}
Custom metrics from EC2 monitoring scripts were visualized in real-time via Amazon CloudWatch dashboards, tracking block latency, RPC success rates, throughput, and resource usage. RPC latency was mostly stable, with brief spikes under load. Alerts for anomalies—like latency less than 1s or delayed inclusion—were rare and resolved without disruption. CloudWatch enabled effective real-time and post-analysis monitoring without needing access to internal AMB infrastructure.

\subsection{Security Analysis}
A multi-layered security evaluation was performed on the deployed system architecture. Access to the AMB Ethereum node was restricted using fine-grained IAM policies, ensuring that only the designated EC2 monitoring instance could interact with the node’s RPC interface. The infrastructure was further protected by deploying all components within a private VPC, with custom security groups and Network Access Control List~\cite{awsNACL} limiting traffic to only essential ports and protocols. TLS encryption was enforced for all RPC traffic~\cite{tlsRFC}, safeguarding data-in-transit from interception or tampering. While AMB abstracts node internals, additional logging and threat detection services—such as AWS CloudTrail~\cite{awsCloudTrail} and GuardDuty~\cite{awsGuardDuty}—can be optionally enabled to support auditability and anomaly detection. These combined safeguards ensure compliance with best practices for cloud-hosted blockchain nodes, protecting against threats such as denial-of-service attacks, credential misuse, and unapproved access attempts. Collectively, the plotted data underscores the viability of Amazon Managed Blockchain for consistent and responsive Ethereum node operations while also highlighting important nuances in gas-based prioritization, transaction throughput, and value transfer dynamics. High-efficiency blocks demonstrate optimized resource usage, while Fig. \ref{fig:tid-gas} reveals how fee strategies affect transaction inclusion delays. Mempool and address-level behavior further validate the effectiveness of our hybrid monitoring setup in capturing real-time blockchain activity. These insights lay the foundation for informed performance tuning, enhanced observability, and secure application deployment on permissioned blockchain platforms.


\section{Limitations and Future Work} \label{sec:limitation_future_work}
While Amazon Managed Blockchain (AMB) offers a secure and scalable foundation for Ethereum node deployment, its managed nature imposes critical constraints for advanced experimentation and blockchain infrastructure research. Key limitations include restricted access to low-level Ethereum client metrics, such as peer connection logs, memory usage, and execution traces, which are vital for diagnosing performance anomalies and studying consensus behavior in detail~\cite{choudhuri2024mempool}. Debug-level RPC endpoints and configurable flags (e.g., `--rpc-apis`, `--syncmode`, `--trace`) are also inaccessible in AMB, impeding protocol instrumentation or consensus modification studies. Additionally, transaction pool (mempool) visibility is constrained to basic RPC queries (e.g., \texttt{eth\_getBlockByNumber}, \texttt{eth\_pendingTransactions}), offering insufficient granularity for capturing real-time transaction propagation, prioritization strategies, or miner-induced reordering~\cite{choudhuri2024mempool}. Although an EC2-based monitoring layer partially restores observability, it introduces polling latency and cannot replicate in-process telemetry achievable in self-hosted nodes. Further limitations include higher operational costs relative to self-hosted setups, limited client diversity (Geth-only access), and regional availability constraints, all of which challenge reproducibility and scalability in latency-critical or large-scale deployments~\cite{eren2025security, ajith2024analyzing}. While we report a detailed breakdown of AMB and EC2 costs, a side-by-side cost comparison with equivalent self-hosted Ethereum nodes was not conducted. Future work will evaluate trade-offs across different deployment models and regions.

To overcome these limitations and enhance research utility, future work will implement a dual-node strategy combining AMB with self-hosted Geth and Besu clients. This hybrid benchmarking setup will allow fine-grained control over execution parameters and support side-by-side analysis across different client behaviors and consensus variants~\cite{eren2025security}. Tools such as Flashbots Explorer and MEV-Inspect will be integrated to provide insight into transaction ordering, miner extractable value (MEV), and inclusion fairness~\cite{flashbots}. Real-time dashboards powered by AWS Amplify and Web3.js will enable dynamic performance tracking and visualization of gas trends, latency metrics, and network congestion. Moreover, we aim to deploy multi-region Ethereum clusters using AWS Transit Gateway to test the resilience and synchronization behavior under geographic distribution. Enhanced security observability will be achieved through the incorporation of AWS GuardDuty, CloudTrail, and Security Hub for detecting anomalies, audit trails, and intrusion events across infrastructure layers~\cite{ajith2024analyzing}. These extensions will position the framework as a robust foundation for both academic inquiry and enterprise-grade blockchain analytics.



\section{Conclusion} \label{sec:conclusion}

This research introduced a novel hybrid, cloud-native architecture for deploying and monitoring Ethereum full nodes using Amazon Managed Blockchain, addressing the trade-off between deployment simplicity and operational transparency. By integrating AMB with EC2-based monitoring, CloudWatch metrics, and infrastructure-as-code via AWS CDK, the framework supports secure, scalable, and research-ready node operations. It enables fine-grained transaction analysis, latency tracking, and anomaly detection, making it suitable for both enterprise applications and academic research. The key novelty lies in its modular, dual-layer design that restores deep observability to a managed environment without compromising ease of deployment. To our best knowledge, this is among the first frameworks to bridge the gap between managed Ethereum services and protocol-level experimentation. Features like custom RPC monitoring and gas-efficiency visualization validate the feasibility of empirical blockchain telemetry on cloud platforms. As Ethereum infrastructure evolves, this architecture offers a reusable foundation for secure, transparent, and scalable deployments. Future work will explore multi-region benchmarking, diversified client analysis, advanced security monitoring, and real-time dashboards for policy-driven insights.




\bibliographystyle{IEEEtran}
\bibliography{bibfile}

\end{document}